\documentclass[11pt,twoside]{article}
\usepackage{asp2010}

\resetcounters

\begin{document}

\title{Study of stellar magnetic activity with CoRoT and {\it Kepler} data}
\author{Savita Mathur$^{1,2}$
\affil{$^1$High Altitude Observatory, NCAR, P.O. Box 3000, Boulder, CO 80307, USA}
\affil{$^2$Laboratoire AIM, CEA/DSM-CNRS-Universit\'e Paris Diderot; IRFU/SAp, Centre de Saclay, 91191 Gif-sur-Yvette Cedex, France}
}

\begin{abstract}
Our knowledge of magnetic activity of stars is mostly based on the study of the Sun and some spectroscopic surveys of a few hundreds of stars. However, the detailed mechanisms of the solar magnetic activity cycle are not fully understood. With the {\it Kepler} mission that is providing exquisite photometric data for hundred of thousands of stars monitored continuously for more than three year, we can study changes in the light curves that can be related to magnetic activity. We show here a few examples of stars for which we did a time-frequency analysis to look for such signature. 
\end{abstract}

\section{Introduction}

Magnetic activity in the stars has been studied for decades. One particular interest is to be able to understand the magnetic activity cycles of our Sun and eventually predict its manifestation in the chromosphere. Many surveys such as in the Mount Wilson that observed emission in Ca II HK or H$_\alpha$ lines have allowed us to study different types of stars in terms of spectral type or age. These studies showed that stars can present regular magnetic cycles, irregular cycles, or even flat behaviours \citep{1995ApJ...438..269B}. Unfortunately, our knowledge of the detailed mechanisms behind the magnetic activity cycles is not complete. For stars with a convection zone, the dynamo theory is believed to rule those cycles but no reliable prediction of the length and strength of the cycles is available so far.

\noindent Some works also showed that short magnetic activity cycles exist. For instance \citet{2009MNRAS.398.1383F} studied eta Boo for which a reversal of polarity was observed with a 1 year periodicity. Besides, \citet{2010ApJ...723L.213M} discovered one of the shortest cycles in iota Hor with a period of 1.6 yr based on Ca II HK measurements. In addition, a magnetic cycle of the same period was measured with Xray observations \citep{2013A&A...553L...6S}. 

\noindent One of the highlights of the CoRoT mission was the first detection of magnetic activity with asteroseismology for the F star HD~49933 \citep{2010Sci...329.1032G}. It had already been observed in the Sun that during the rise of the cycle, the frequencies of the acoustic modes increase and their amplitudes decrease \citep{2009A&A...504L...1S}. Unfortunately, some further studies of three additional CoRoT solar-like targets did not lead to a firm detection of a magnetic activity signature  \citep{2013A&A...550A..32M}.

\noindent The {\it Kepler} mission has been observing the Cygnus constellation field for more than 3.5 years now. These unprecedented data already allowed us to study the internal structure and rotation of solar-like stars \citep{2012ApJ...756...19D,2012ApJ...749..152M} and red giants \citep{2011Natur.471..608B,2012Natur.481...55B,2012A&A...548A..10M}. Since the launch of the mission, many solar-like stars have been observed for many months and have been analysed \citep[e.g.][]{2011A&A...534A...6C,2011ApJ...733...95M}. Now we have this great opportunity to study the magnetism of the stars thanks to the long and continuous data of the mission.

\section{Looking for hints of magnetic activity}

We will show here the analysis of two solar-like stars KIC~11026764 \citep{2010ApJ...723.1583M} and KIC~11395018 \citep{2011ApJ...733...95M} that present very different behaviours in their light curves.

\subsection{Data and methodology}

The data used for this work have been calibrated following the method described by \citet{2011MNRAS.414L...6G}. Since we are not interested in the acoustic modes propagating in these stars, we are using the long cadence data ($\delta t \sim$\,29.58\,min), which gives us the opportunity to analyse more than 1200 days of data.  

\noindent We applied the wavelet techniques \citep{1998BAMS...79...61T,liu2007,2010A&A...511A..46M} to do a time-frequency analysis of these two stars. This has for instance been done on some CoRoT targets to look for the surface rotation of the star. It basically consists of looking at the correlation between the light curve and a mother wavelet (the Morlet wavelet) for which we change the period and that we slide along the time series. This has been tested on a large number of stars observed by CoRoT and {\it Kepler} \citep[e.g.][]{2010A&A...518A..53M}. This is a very interesting tool to see when the active regions are more predominant in the light curve and gives the signature of the rotation period of the star. We compute the magnetic proxy by projecting the wavelet power spectrum on the time axis around the rotation period of the star. \citet{2013arXiv1301.6930G} applied this tool to 16 years of VIRGO data and showed that this projection is a good proxy for magnetic activity after comparing it with the radio flux at 10.7cm. 

\subsection{Some examples}

We present here the analysis of the two solar like stars KIC~11026764 and KIC~11395018. These stars were among the first ones that were seismically analysed after the launch of the {\it Kepler} mission. They are rather evolved solar-like stars that present avoided crossings due to the presence of mixed modes. 

\begin{figure}[htbp]
\begin{center}
\includegraphics[width=6.5cm, trim=0cm 0 2.5cm 0]{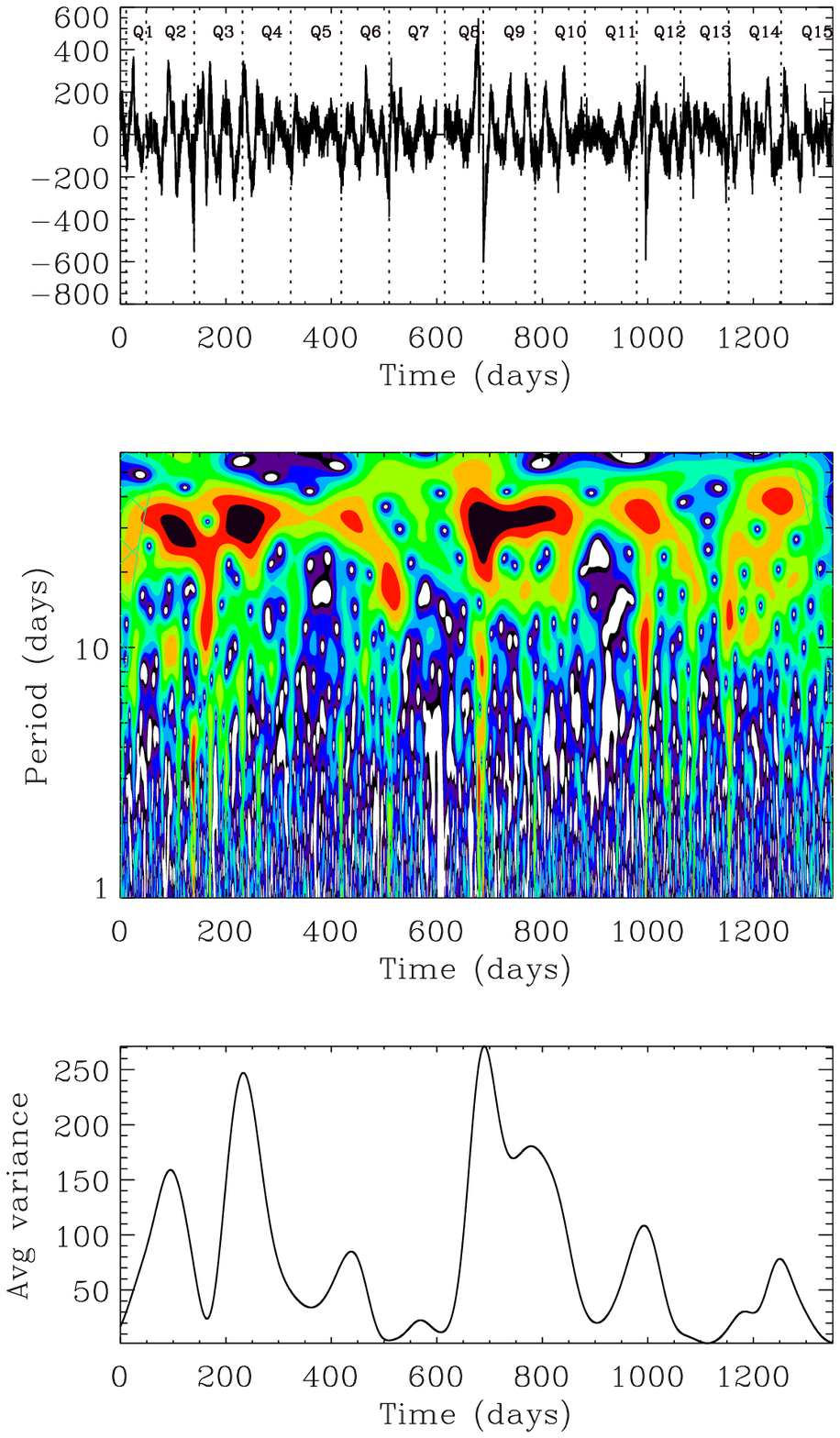}
\includegraphics[width=6.5cm, trim=0cm 0 2.5cm 0]{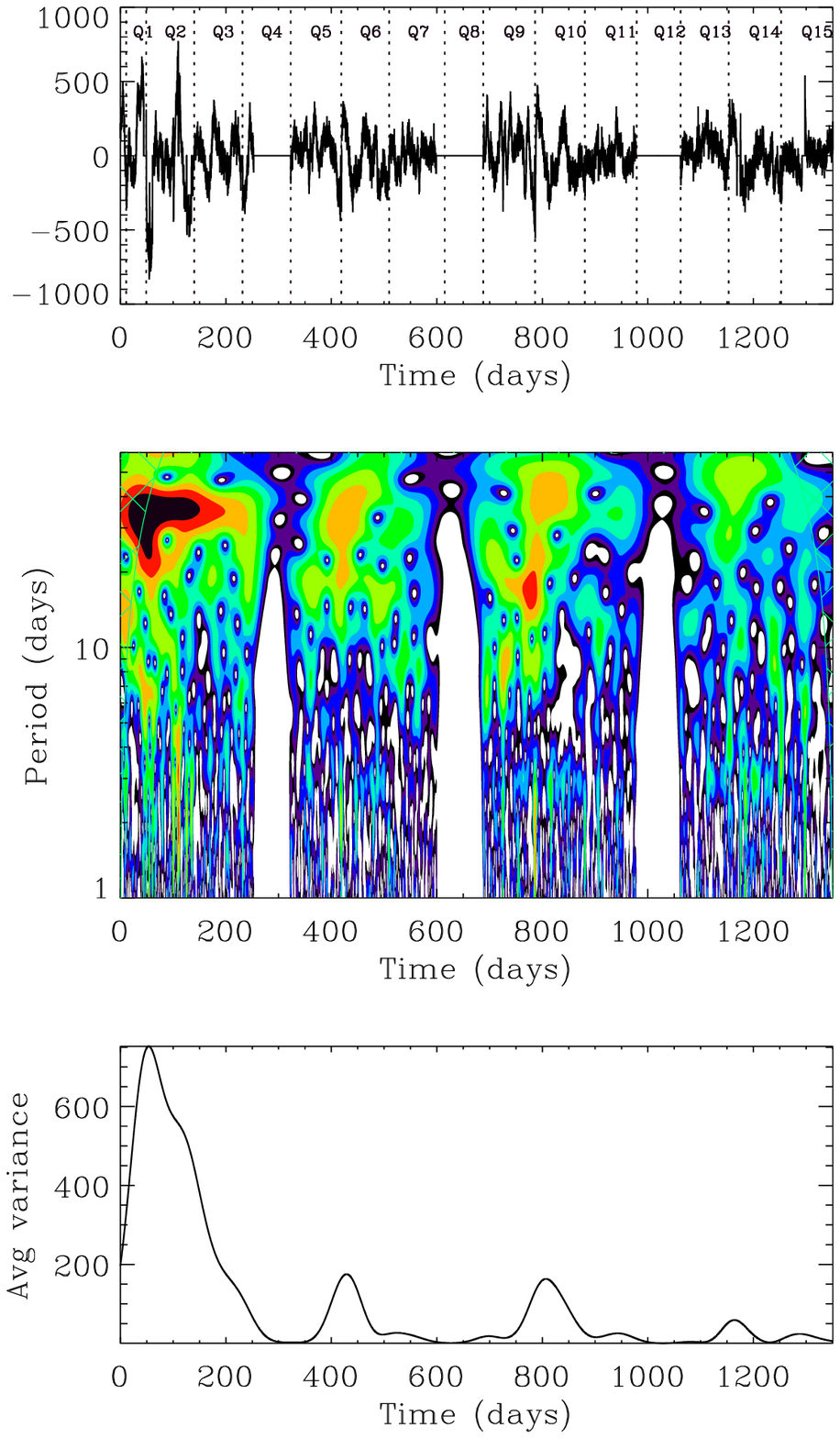}
\caption{Wavelet analysis of more $\sim$\,1200 days of long-cadence data obtained by {\it Kepler} for: KIC~11026764 (left panel) and KIC~11395018 (right panel). Top panel shows the light curve with the dot lines delimiting the quarters. The middle panel is the wavelet power spectrum as a function of period and time obtained as described in Section 2.1. Red and dark (resp. blue and purple) colours correspond to high (resp. low)  power. Bottom panel is the average variance or magnetic proxy computed around the rotation period of the star: between 30 and 37 days for KIC~11026764 and between 25 and 40 days for KIC11395018. }
\label{fig1}
\end{center}
\end{figure}

\noindent \citet{2010ApJ...723.1583M} studied KIC~11026764 with a deep asteroseismic analysis based on the individual frequencies of the acoustic modes of the star. This study was done with only one month of {\it Kepler} data. They found a mass for the star between 1.13\,$M_{\odot}$ and 1.23\,$M_{\odot}$ and an age of 5.5\,Gyr.  The wavelet analysis (left panel of Figure~\ref{fig1}) allows us to infer a rotation period of 32.8 days with continuous high power around that period in the wavelet power spectrum. We notice that the power is higher between 200 and 300 days and between 600 and 900 days. This can also be seen in the magnetic proxy. This suggests that there are some persistent active regions present on the surface of the star. After 900 days, the magnetic proxy suggests a lower magnetic level. 


\noindent Another interesting star is KIC~11395018, which was modelled by both grid-modeling \citet{2012A&A...537A.111C} and an asteroseismic analysis using the individual mode frequencies \citep{2013ApJ...763...49D}. Their results agree within the uncertainties. The final parameters of the detailed modelling are the following: M\,=\,1.27\,$\pm$\,0.04\,$M_{\odot}$, and $\tau$\,=\,4.57\,$\pm$\,0.23Gyr. The wavelet analysis is shown in Figure~\ref{fig1} (right panel). Unfortunately this star located on the CCD that was on the module that broke in 2010 January, which is responsible for the 3-month gaps every year. However, the wavelet power spectrum suggests a rotation period of 35.3 days for this star. We notice that the power around that period in the wavelet power spectrum decreases with time, which is also seen in the magnetic proxy. With this observation, we can assume that we might be observing the star during the decrease of its magnetic activity.



\section{Conclusions}

The time-frequency analysis of two solar-like stars for which we know the fundamental properties thanks to asteroseismology allows us to extract the surface rotation period of the stars based on the presence of active regions on their surface. Based on this information, we can start to study the magnetic activity of the stars around the rotation period found. For KIC~11026764, which has a surface rotation period of $\sim$\,33 days, we do not see any clear modulation of a cycle but we see changes in the magnetic proxy related to the magnetic activity of the star. For the other star KIC~11395018, which has a rotation period of $\sim$\,35 days, we observe a slight decrease of the magnetic proxy, suggesting a decrease of its magnetic activity cycle. Longer datasets would be needed for these rather slow rotators to observe a complete magnetic activity cycle. A similar analysis is being done on more than 20 stars that rotate faster than 12 days  \citep{mathur2013}.

\acknowledgements NCAR is partially funded by the National Science Foundation. SM acknowledges the support of the European Community's Seventh Framework Programme (FP7/2007-2013) under grant agreement no. 269194 (IRSES/\-ASK), of the CNES, and of the University of Tokyo. SM thanks R.~A. Garc\'ia and J. Ballot for useful comments and discussion.

\bibliographystyle{asp2010} 
\bibliography{/Users/Savita/Documents/BIBLIO_sav.bib}

\end{document}